\documentclass[twocolumn,showpacs,preprintnumbers,amsmath,amssymb,10pt]{revtex4}
\usepackage{graphicx}% Include figure files
\usepackage{dcolumn}% Align table columns on decimal point
\usepackage{bm}% bold math
\usepackage{cancel} % allows the use of cancellation of terms.
\begin{document}
%\preprint{APS/123-QED}

\title{Origin of Intrinsic Gilbert Damping }% Force line breaks with \\
\author{M. C. Hickey} \altaffiliation[Electronic mail : ]{hickey@mit.edu}%Lines break automatically or can be forced with \\
\affiliation{%
Francis Bitter Magnet Laboratory, Massachusetts Institute of Technology, 150 Albany Street, Cambridge, Massachusetts 02139 USA.
}%

\author{J. S. Moodera} %Lines break automatically or can be forced with \\
\affiliation{%
Francis Bitter Magnet Laboratory, Massachusetts Institute of Technology, 150 Albany Street, Cambridge, Massachusetts 02139 USA.
}%
\begin{abstract}
The damping of magnetization, represented by the rate at which it relaxes to equilibrium, is successfully
modeled as a phenomenological extension in the Landau-Lifschitz-Gilbert equation. This is the damping torque term known as Gilbert damping and its direction is given by the vector product of the magnetization and its time derivative. Here we derive the Gilbert term from first principles by a non-relativistic expansion of the Dirac equation. We find that this term arises when one calculates the time evolution of the spin observable in the presence of the full spin-orbital coupling terms, while recognizing the relationship between the curl of the electric field and the time varying magnetic induction.
\end{abstract}
\maketitle

The Gilbert damping torque in magnetic systems describes the relaxation of magnetization and it was introduced into the Laudau-Lifschitz equation \cite{LL,laudau_lifschitz} for describing spin dynamics. Gilbert damping is understood to be a non-linear spin relaxation phenomenon and it controls the rate at which magnetization spins reach equilibrium. The introduction of this term is phenomenological in nature \cite{1353448} and the question of whether it has an intrinsic physical origin has not been fully addressed, in the face of rather successful modeling of the relaxation dynamics of measured systems. Correlating ferromagnetic resonance spectral line-widths \cite{patton:1358,JJAP.45.3889} in magnetic thin films with the change in damping has been successful for confirming the form of
the damping term in the underlying dynamical equations. The intrinsic origin of
the damping itself is still an open question. The damping constant, $\alpha$ is often reformulated in terms of a relaxation time, and
the dominant relaxation processes are invoked to calculate this, but this approach presupposes precessional damping torque. \\
 It has been long thought that intrinsic Gilbert damping had its origin in spin-orbital coupling because this mechanism does not conserve spin, but it has never been derived from a coherent framework.
Non-local spin relaxation processes \cite{tserkovnyak:1375} and disorder broadening couple to the spin dynamics and can enhance the Gilbert damping extrinsically in thin films and heterostructures. This type of spin relaxation, which
is equivalent to ensemble dephasing \cite{tserkovnyak:5234}, is modeled as the (S-S$_{0})$/T$^{*}_{2}$ decay term in the dynamical Bloch equation, where T$^{*}_{2}$ is the decay time of the ensemble of spins. Crudely speaking, during spin relaxation, some spins lag behind the mean magnetization vector and the exchange and magnetostatic fields then exert a time dependent torque.
%Magnon and two magnon dissipation, time dependent creation of electron and hole spin states and spin-orbital coupling have been shown to contribute to the Gilbert damping.
Calculations on relaxation driven damping of this kind presuppose the Gilbert damping term itself which begs the question.\\
The inhomogeneous damping term can be written as ${\bf M}\times d \nabla^{2} {\bf M}/d t$ which gives rise to non-local effects such as spin wave dissipation \cite{tserkovnyak:1375,eilers:054411}. These non-local theories are successful in quantifying the enhancement of the Gilbert damping, but do not derive the intrinsic Gilbert term itself.
There are models \cite{KamberskyV_Can,steiauf:064450} which deal with the scattering of electron spins from thermal equilibrium in the presence of
phonon and spin-orbital interactions which is a dynamic interaction and this allows us to determine the strength of the Gilbert damping for itinerant ferromagnetic metals, generalizing the Gilbert damping response to a tensorial description.  Both the s-d exchange relaxation models \cite{PhysRevLett.93.127204,fahnle:172408} and the Fermi surface breathing models of Kambersky \cite{KamberskyV_Can,PhysRevB.65.212411} either presuppose a Gilbert damping term in the dynamical equation or specify
a phenomenological Hamiltonian H = -1/($\gamma M_{s}$){\boldsymbol $\hat{\alpha}$}.d{\bf M}/dt.
While this method is {\it ab initio} from the point of view of
electronic structure, it already assumes the Gilbert term ansatz.
Hankiewicz {\it et al.} \cite{hankiewicz:020404} construct the inhomogeneous Gilbert damping by connecting the spin density-spin current conservation law with the imaginary part of magnetic susceptibility tensor and show that both electron-electron and impurity scattering can enhance the damping through the transverse spin conductivity for finite wavelength excitations (${\bf q} \neq 0$).
In previous work \cite{hankiewicz:174434}, there are derivations of the Gilbert constant by comparing
the macroscopic damping term with the torque-torque correlations in homogeneously magnetized electron gases possessing spin orbital coupling.
For the case of intrinsic, homogeneous Gilbert damping, it is thought that in the absence of spin-orbital scattering, the damping vanishes. We aim to focus on intrinsic, homogeneous damping and its physical origin in a first-principles framework and the question as to whether spin in a homogeneous time-varying magnetization can undergo Gilbert damping is addressed.\\
In this work, we show that Gilbert damping does indeed arise from spin-orbital coupling, in the sense that
 it is due to relativistic corrections to the Hamiltonian which couple the spin to the electric field and we arrive at the Gilbert damping term by first writing down the Dirac equation for electrons in magnetic and electric potentials. We transform the Hamiltonian in such a way as to write it in a basis in which the canonical momentum terms are even powers. This is a standard approach in relativistic quantum mechanics and we do this in order to calculate the terms which couple the linear momentum to the spin in a basis which is diagonal in spin space. This is often referred to as a non-relativistic expansion of the Dirac equation. This allows us to formulate the contributions as a perturbation to an otherwise non-relativistic particle. We then wish to calculate the rate equation for the spin observable with all of the spin-orbital corrections in mind. \\
Now, we start with a purely relativistic particle, a Dirac particle and we write the Dirac-Pauli Hamiltonian, as follows :
\begin{align}
H &= c{\boldsymbol \alpha}.({\bf p} - e \frac{{\bf A}}{c}) + \beta m_{0} c^{2}+e \phi\\
  &=  \mathcal{O} +\beta m_{0}c^{2}  +\varepsilon
\end{align}
where {\bf A} and $\phi$ are the magnetic vector potential and the electrostatic potential, respectively and
\begin{displaymath}
{\boldsymbol \alpha}  = \left(
                                      \begin{array}{cc}
                                        0 & \sigma_{i} \\
                                        \sigma_{i} & 0 \\
                                      \end{array}
                                    \right)
\end{displaymath}
while
\begin{displaymath}
\beta = \left(\begin{array}{cc}
                                                              1 & 0 \\
                                                              0 & -1
                                                            \end{array}\right).
\end{displaymath}
We observe immediately that
$\beta\mathcal{O} = -\mathcal{O}\beta$. $\mathcal{O}$ is the Dirac canonical momentum , c and e are the speed of light in a vacuum and the electronic charge, respectively.\\
We now need to rewrite the Hamiltonian in a basis where the odd operators (whose generators are off diagonal in the Pauli-Dirac basis : $\alpha^{i}$, $\gamma^{i}$,$\gamma_{5}$ ..)
and even operators (whose generators are diagonal in the Pauli-Dirac basis : ({\bf 1}, $\beta$, $\Sigma$,.. ) are decoupled from one another. \\
If we are to find S so that H$^{'}$ does not contain odd powers of spin operators, we must chose the operator S, in such a way as to satisfy the following constraint :
\begin{eqnarray}
[S,\beta] = \frac{-\mathcal{O}}{im_{0} c^{2}}
\end{eqnarray}
In order to satisfy cancelation of the odd terms of $\mathcal{O}$ to first order, we require $S =\frac{-i\mathcal{O}\beta}{2 m_{0} c^{2}}$ and this is known as the Foldy-Wouthuysen transformation in relativistic quantum mechanics and it is treated in some detail in, for example, reference
\cite{W.Greiner}. We now would like to collect all of the terms into the transformed Hamiltonian, and this is written as
\begin{align*}
H^{'} = \beta\left(m_{0}c^{2} + \frac{\mathcal{O}^{2}}{2m_{0}c^{2}}-\frac{\mathcal{O}^{4}}{8m_{0}^{3}c^{6}}\right)\\
+\mathcal{\varepsilon}-\frac{1}{8m_{0}^{2}c^{4}}[\mathcal{O},[\mathcal{O},\mathcal{\varepsilon}]]%-\frac{i}{8m_{0}^{2}c^{4}}[\mathcal{O},\dot{\mathcal{O}}]\\
+\frac{\beta}{2m_{0}c^{2}}[\mathcal{O},\mathcal{\varepsilon}]-\frac{O^{3}}{3m_{0}^{2}c^{4}}%+\frac{i\beta\dot{\mathcal{O}}}{2m_{0}c^{2}} = \beta m c^{2} + \mathcal{\varepsilon}^{'} + \mathcal{O}^{'}
\label{expansion_4}
\end{align*}
The expression above contains odd powers of the canonical momentum $\mathcal{O}$, so we redefine the canonical momentum to encapsulate all of these odd power terms. So we now apply the procedure of eliminating odd powers once again :
\begin{equation}
S^{'} = -\frac{i\beta}{2m_{0}c^{2}}\mathcal{O^{'}} = \frac{-i\beta}{2m_{0}c^{2}}\left(\frac{\beta}{2m_{0}c^{2}}[\mathcal{O},\mathcal{\varepsilon}]-\frac{\mathcal{O}^{3}}{3m_{0}^{2}c^{4}}\right)%+\frac{i\beta\dot{\mathcal{O}}}{2m_{0}c^{2}}\right)
\end{equation}
\begin{equation}
H^{''} = e^{iS^{'}}H^{'}e^{-iS^{'}} = \beta m_{0}c^{2}+\mathcal{\varepsilon^{'}}+\mathcal{O^{''}},
\end{equation}
where $\mathcal{O^{''}}$ is now O($\frac{1}{m_{0}^{2}c^{4}}$), which can be further eliminated by applying
a third transformation (S$^{''}$ = $\frac{-i\beta\mathcal{O^{''}}}{2m_{0}c^{4}}$), we arrive at the following
Hamiltonian :
\begin{align*}
H^{'''} &= e^{iS^{''}}\left(H^{''}\right)e^{-iS^{''}} = \beta m_{0}c^{2}+\mathcal{\varepsilon^{'}}\\
&=\beta\left(m_{0}c^{2}+\frac{\mathcal{O}^{2}}{2m_{0}c^{2}}-\frac{\mathcal{O}^{4}}{8m_{0}^{3}c^{6}}\right)+\\
&\mathcal{\varepsilon}-\frac{1}{8m_{0}^{2}c^{4}}[\mathcal{O},[\mathcal{O},\mathcal{\varepsilon}]]
\label{expansion_5}
\end{align*}
Thus we have the fully Foldy-Wouthuysen transformed Hamiltonian :
\begin{align*}
&H^{'''} = \beta\left(m_{0}c^{2}+\frac{({\bf p}-e{\bf A}/c)^{2}}{2 m_{0}} -\frac{{\bf p}^{4}}{8m_{0}^{3}c^{6}}\right)+e\Phi\\
&-\frac{e\hbar}{2m_{0}c^{2}}\beta{\boldsymbol \Sigma}.{\bf B}-\frac{ie\hbar^{2}}{8m_{0}^2c^2}{\boldsymbol \Sigma}.(\nabla\times{\bf E})\\
&-\frac{e\hbar}{4m_{0}^2c^{2}}{\boldsymbol \Sigma}.{\bf E}\times{\bf p}-\frac{e\hbar^{2}}{8m_{0}^2c^{2}}(\nabla.{\bf E})
\end{align*}
The terms which are present in the above Hamiltonian, show us that we have a ${\bf p}^{4}$ kinetic part which
is the relativistic expansion of the mass of the particle.  The terms which couple to the
spin {\bf $\Sigma$} are of importance and we see that these terms correspond to the Zeeman, spin-orbital (comprising momentum and electric field curl terms) and the Darwin term, respectively. Strictly speaking, the presence of the scalar
 potential $\phi$ breaks the gauge invariance in the Pauli-Dirac Hamiltonian and a fully gauge invariant theory would require that this contain the gauge-free electromagnetic field energy. We omit the term $\frac{e^2\hbar}{4m^{2}c^{3}}{\boldsymbol \Sigma}.({\bf A}\times {\bf E})$ (which establishes gauge invariance in the momentum terms) in this rotated Hamiltonian, as it is O($1/m^{2}c^{3}$) and we are only interested in calculating semiclassical rate equations for fields, which are manifestly gauge-invariant, and not wavefunctions or energy eigenvalues. We can now define the spin dependent corrections to a non-relativistic Hamiltonian :

\begin{equation}
H^{\boldsymbol \Sigma} = -\frac{e\hbar}{2m_{0}c^{2}}\beta{\boldsymbol \Sigma}.{\bf B} - \frac{e\hbar}{4m_{0}^{2}c^{2}}{\boldsymbol \Sigma}.{\bf E}\times{\bf p} -\frac{ie\hbar^{2}}{8m_{0}^{2}c^{2}}{\boldsymbol \Sigma}.(\nabla\times{\bf E}).
\label{HSO}
\end{equation}
where
\begin{displaymath}
{\bf \Sigma } = \left(
                         \begin{array}{cc}
                           {\boldsymbol \sigma}_{i} & 0 \\
                           0 & {\boldsymbol \sigma}_{i} \\
                         \end{array}
                       \right)\equiv{\bf \hat{S}}_{i}.
\end{displaymath}
and $\sigma_{i}$ are the Pauli matrices. Note that the last two terms in Equation \ref{HSO} encapsulate the
entire spin orbital coupling in the sense that these terms couple the particle's linear momentum to the spin  ${\bf \hat{S}}_{i}$. The first spin-orbital term in the Hamiltonian is well known and give rise to momentum dependent magnetic fields. When the ensuing dynamics are calculated for this case, it gives rise to spin relaxation terms which are linear in spin \cite{engel:036602}. Note that, while neither spin-orbital term is Hermitian, the two terms taken together {\it are } Hermitian and
so the particles angular momentum is a conserved quantity and the total energy lost in going from collective spin excitations (spin waves) to single particles states via spin-orbital coupling is gained by the electromagnetic field. Recognizing the curl of the electric field in the last term, we
now rewrite this the time varying magnetic field as given by Maxwells equations as $\nabla\times{\bf E}=-\frac{\partial {\bf B}}{\partial t}$. We now have an explicitly time-dependent perturbation on the non-relativistic Hamiltonian.
We can write the time-varying magnetic field seen by the spin (in, for example a magnetic material) as $\frac{\partial {\bf B}}{\partial t} =\frac{\partial {\bf B}}{\partial {\bf M}}\cdot\frac{\partial {\bf M}}{\partial t} = \mu_{0}(1+\chi_{m}^{-1})\frac{\partial {\bf M}}{\partial t}$. We now have the spin dependent Hamiltonian : \\
\begin{align*}
H^{{\bf S}}= -\frac{e\hbar}{2m_{0}c^{2}}\beta{\bf S}.{\bf B} - \frac{e\hbar}{4m_{0}^{2}c^{2}}{\bf S}.{\bf E}\times{\bf p}& \\
+i\frac{e\hbar^{2}\mu_{0}}{8m_{0}^{2}c^{2}}{\bf S}.\left(1+\chi_{m}^{-1}\right).\frac{d{\bf M}}{dt}&=\\
H^{{\bf S}} = H_{0}^{\bf S}+H^{\bf S}(t).
\end{align*}
We focus our attention on the explicitly time-dependent part of the Hamiltonian H$^{\bf S}$(t) ;
\begin{equation}
 H^{{\bf S}} (t) = i\frac{e\hbar^{2}\mu_{0}}{8m_{0}^{2}c^{2}}{\bf S}.\left(1+\chi_{m}^{-1}\right).\frac{d{\bf M}}{dt}.
 \label{Gilbert_Hamiltonian}
 \end{equation}
%In order to investigate the time evolution of the spin observable with this Hamiltonian, we invoke the Quantum-Liouville
%description of the the time-evolution operator (U(t)). This can be written as follows :
%\begin{equation}
%U(t) = \displaystyle \overrightarrow{T}e^{\frac{-i}{\hbar}\int_{t_{0}}^{t}H(t^{'})dt^{'}},
%\end{equation}
%where $\overrightarrow{T}$ is the time-ordering.

%The time dependent spin observable {\bf S}(t) can be written as :

%\begin{align*}
%{\bf S}(t) = \langle\Psi_{0}(t_{0})|e^{\frac{i}{\hbar}\int_{t_{0}}^{t}H(t^{'})dt^{'}}\overleftarrow{T}S\overrightarrow{T}e^{\frac{-i}{\hbar}\int_{t_{0}}^{t}H(t^{'})dt^{'}}|\Psi_{0}(t_{0})\rangle
%\end{align*}.

In this perturbation scheme, we allow the Hermitian components of the Hamiltonian to define the ground sate of the system and we treat the explicitly time-dependent Hamiltonian (containing the spin orbital terms) as a
time dependent perturbation. In this way, the rate equation is established from a time dependent perturbation expansion in the quantum Liouville description.
We now define the magnetization observable as ${\bf \hat{M}}$ =$\displaystyle\sum_{\alpha}\frac{g\mu_{B}}{V}Tr\rho{\bf \hat{S}}^{\alpha}(t)$ where the summation is taken over the site of the magnetization spin $\alpha$. We now examine the
time dependence of this observable by calculating the rate  equation according to the quantum-Liouville rate equation ;
\begin{align}
&\frac{d \rho(t)}{dt}+ \frac{1}{i\hbar}[\hat\rho,H] = 0
\label{rho_equation}
\end{align}
This rate equation governs the time-evolution of the magnetization observable as defined above, in the non-equilibrium regime. We can write the time derivative of the magnetization \cite{PhysRevLett.92.097601}, as follows ;
\begin{align*}
\frac{d {\bf M}}{d t} = \displaystyle\sum_{n,\alpha}\frac{g\mu_{b}}{V}\langle\Psi_{n}(t)|\frac{1}{i\hbar}[\rho{\bf S}^{\alpha},H]+\frac{\partial\rho}{\partial t}{\bf {\bf S}^{\alpha}}+\rho\frac{\partial {\bf S}^{\alpha}}{\partial t}|\Psi_{n}(t)\rangle,\\
\end{align*}
and we can use the quantum Liouville rate equation as defined by Equation \ref{rho_equation} to simplify this expression and we arrive at the following rate equation :
\begin{align}
&\frac{d {\bf M}}{d t}  = \displaystyle\sum_{\alpha}\frac{g\mu_{b}}{V}\frac{1}{i\hbar}Tr\{\rho[{\bf S}^{\alpha},H^{\bf S}(t)]\}
\end{align}
In the case of the time dependent Hamiltonian derived in equation \ref{Gilbert_Hamiltonian}, we can assume a first order dynamical
 equation of motion given by $\frac{d{\bf M}}{dt} = \gamma {\bf M}\times {\bf H}$ and calculate the time evolution for the magnetization observable :
\begin{align*}
&\frac{d {\bf M}}{d t}  = \displaystyle\sum_{\alpha,\beta}\frac{g \mu_{B}}{V}\frac{1}{i\hbar}Tr\rho[S^{i}_{\alpha},\frac{ie\hbar^{2}\mu_{0}}{8m^{2}c^{2}} S^{j}_{\beta}].(1+\chi_{m}^{-1})\frac{\overleftrightarrow{\partial}{\bf M}}{d t}\\
%&\frac{d {\bf M}}{d t}  = \displaystyle\sum_{\alpha,\beta}\frac{g \mu_{B}}{V}\frac{ie\hbar^{2}\mu_{0}}{8m^{2}c^{2}}\frac{1}{\hbar}Tr\rho[S^{i}_{\alpha},S^{j}_{\beta}](1+\chi_{m}^{-1})\delta_{jl}\frac{\overleftrightarrow{\partial} M_{l}}{d t}\\
&  = \displaystyle\sum_{\alpha}\frac{g \mu_{B}}{V}\frac{ie\hbar^{2}\mu_{0}}{8m^{2}c^{2}}\frac{1}{i\hbar}Tr\rho i\hbar\epsilon_{ijk}S^{k}_{\alpha}\delta_{\alpha\beta}(1+\chi_{m}^{-1})\delta_{jl}\frac{\overleftrightarrow{\partial} M_{l}}{d t}\\
&=-i\frac{e\hbar\mu_{0}}{8m^{2}c^{2}} (1+\chi_{m}^{-1}){\bf M}\times\frac{\overleftrightarrow{\partial} {\bf M}}{d t},\\
\end{align*}
where, in the last two steps, we have used the following commutation relations for magnetization spins :
$[S^{i}_{\alpha},S^{j}_{\beta}] =i\hbar\epsilon_{ijk}S^{k}_{\alpha}\delta_{\alpha\beta}$ which implies that the theory presented here is that which relates to local dynamics and that the
origin of the damping is intrinsic. We now recognize the last equation as the which describes Gilbert damping, as follows :
\begin{align}
&\frac{d {\bf M}}{d t}  = -\frac{\alpha}{M_{s}}.{\bf M}\times\frac{\overleftrightarrow{\partial} {\bf M}}{\partial t}
\end{align}
whereby the constant $\alpha$ is defined as follows :

\begin{align}
&\alpha = \frac{ie\hbar\mu_{0}M_{s}}{8m_{0}^{2}c^{2}}\left(1+\chi_{m}^{-1}\right)
\label{alpha_eq}
\end{align}

The $\alpha$ defined above corresponds with the Gilbert damping found in the phenomenological term
in the Landau-Lifschitz-Gilbert equation and $\chi_{m}$ is the magnetic susceptibility. In general, the inverse of the susceptibility can be written in the form \cite{PhysRevLett.88.056404},
\begin{equation}
\chi^{-1}_{ij}({\bf q},\omega)=\tilde{\chi}^{-1}_{\bot}
({\bf q},\omega)-\frac{\omega_{ex}}{\gamma \mu_{0}M_{0}}\delta_{ij},
\end{equation}
where the equilibrium magnetization points along the z-axis and
$\omega_{ex}$ is the excitation frequency associated with the internal exchange field.
The $\delta_{ij}$ term in the inverse susceptibility does not contribute
to damping mechanisms as it corresponds to the equilibrium response.
 In the basis (M$_{x}$$\pm$ iM$_{y}$,M$_z$), we have the dimensionless transverse
magnetic susceptibility, as follows :
 \begin{align*}
\tilde{\chi}_{m\bot}({\bf q},{\bf \omega}) = \gamma\mu_{0}\frac{M_{0}-i\gamma\sigma_{\bot}q^2}{\omega_0-\omega-i\gamma\sigma_{\bot}q^2\omega_0/M_0}
\end{align*}

\begin{figure}[ht]
\begin{center}
 \includegraphics[width=3.0in]{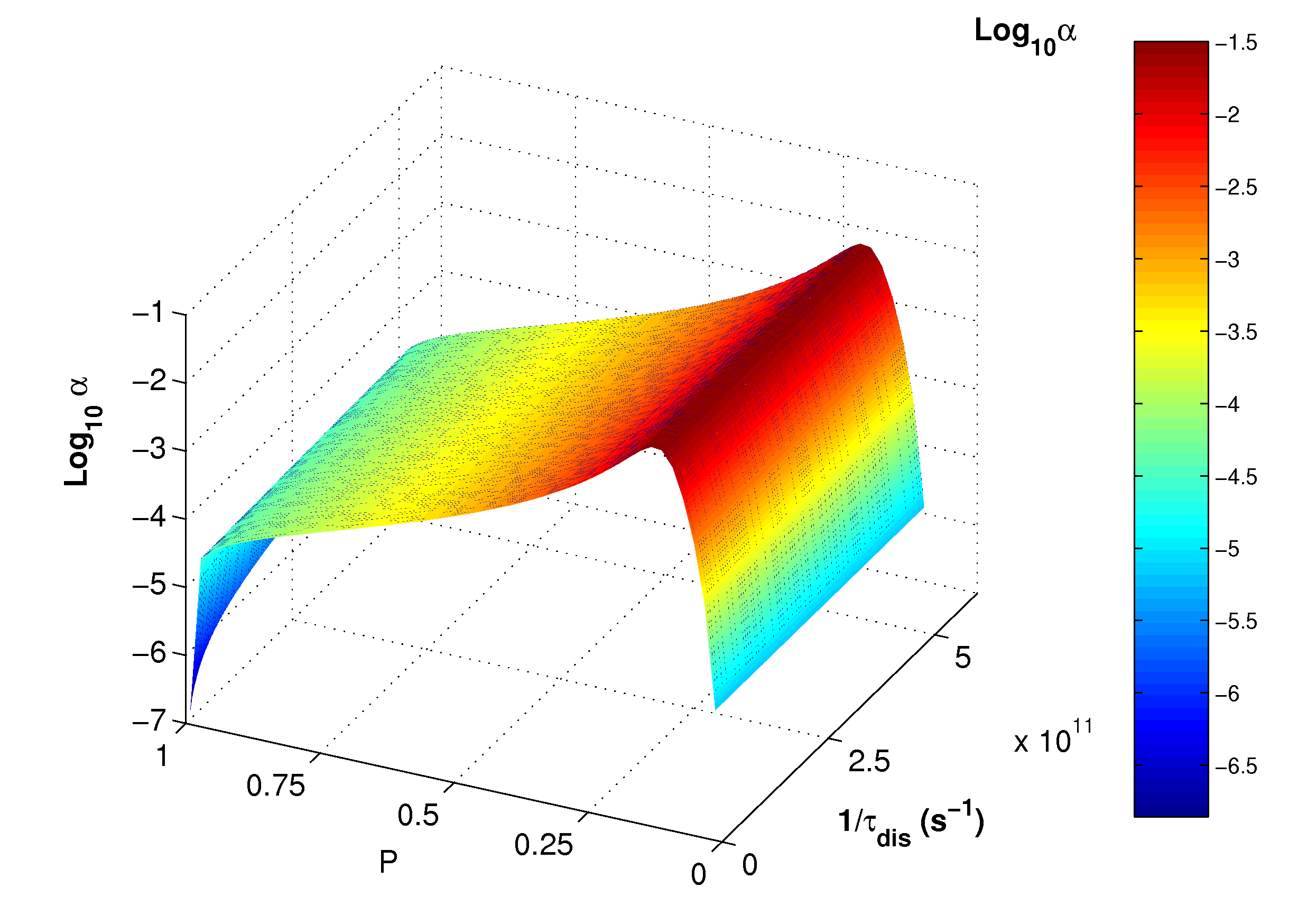}
\end{center}
 \caption{(Color Online) Plot of the dimensionless Gilbert damping constant $\alpha$ in the dc limit ($\omega$=0),
 as a function of electron spin polarization and disorder scattering rate.} \label{Fig1}
\end{figure}

 The first term in the dimensionless Gilbert coefficient (Equation \ref{alpha_eq}) is small ($\sim$ 10$^{-11}$) and the higher damping rate is controlled by the
the inverse of the susceptibility tensor. For uniformly saturated magnetization, the damping is critical and so the system is already at equilibrium as far as the Gilbert mechanism is concerned (d{\bf M}/dt = 0 in this scenario).
The expression for the dimensionless damping constant $\alpha$  in the dc limit ($\omega$ =0 ) is :
\begin{align}
\alpha = \frac{e\hbar\mu_{0}M_{s}}{8m^2_{0}c^2}Im\left(\frac{\frac{\omega_0}{\gamma \mu_{0} M_{0}}-\frac{i\sigma_{\bot}{\bf q}^{2}\omega_{0}}{\mu_{0}M_{0}^{2}}}{1-i\gamma\sigma_{\bot}{\bf q}^{2}/M_{0}}\right),
\end{align}

and we have the transverse spin conductivity from the following relation (in units whereby $\hbar$=1) :
\begin{align*}
\sigma_{\bot}=\frac{n}{4m^{*}\omega_0^{2}}\left(\frac{1}{\tau_{\bot}^{dis}}+\frac{1}{\tau_{\bot}^{ee}}\right),
\end{align*}
where $\tau_{\bot}^{dis}$ and $\tau_{\bot}^{ee}$ are the impurity disorder and electron electron-electron scattering times as defined  and parameterized in Reference \cite{hankiewicz:020404}.
We calculate the extrinsically enhanced Gilbert damping using the following set of parameters as defined in the same reference ; number density of the electron gas, n=1.4$\times$10$^{27}$ m$^{-3}$, polarization p, equilibrium magnetization M$_0$= $\gamma$pn/2, equilibrium excitation frequency $\omega_0=E_{F}[(1+p)^{2/3}-(1-p)^{2/3}]$ and wave-number defined as q = 0.1 k$_{F}$, where E$_{F}$  and k$_{F}$ are the Fermi energy and Fermi wave number, respectively. m$^{*}$ is taken to be the electronic mass. Using these quantities, we evaluate $\alpha$ values and these are plotted as a function of both polarization and disorder scattering rate in Figure \ref{Fig1}.\\
 In general, the inverse susceptibility $\chi_{m}^{-1}$ will determine the strength of the damping in real inhomogeneous magnetic systems where spin relaxation takes place, sub-bands are populated by spin orbit scattering and spin waves and spin currents are emitted. The susceptibility term gives the Gilbert damping a tensorial quality, agreeing with the analysis in Reference \cite{steiauf:064450}.
 Further, the connection between the magnetization dynamics and the electric field curl provides the mechanism for the energy loss to the electromagnetic field. The generation of radiation is caused by
the rotational spin motion analog of electric charge acceleration and the radiation spin interaction term has the form :
\begin{align}
 H^{{\bf S}} (t) = i\frac{e\hbar^{2}\mu_{0}}{8m_{0}^{2}c^{2}}\displaystyle\sum_{\alpha}\left(1+\chi_{m}^{-1}\right) {\bf S}_{\alpha}.\frac{d{\bf M}}{dt}.
 \end{align}
In conclusion, we have shown that the Gilbert term, heretofore phenomenologically used to describe
damping in magnetization dynamics, is derivable from first principles and its origin lies in spin-orbital coupling. By a non-relativistic expansion of the Dirac equation, we show that there is a term which contains the curl of the
electric field. By connecting this term with Maxwells equation to give the total time-varying magnetic induction, we have found that this damping term can be deduced from the rate equation for the spin observable, giving the correct
vector product form and sign of Gilberts' original phenomenological model. Crucially, the connection of the time-varying
magnetic induction and the curl of the electric field via the Maxwell relation shows that the damping of magnetization dynamics is commensurate with the emission of electromagnetic radiation and the radiation-spin interaction is specified from first principles arguments.

%The dimensions of the second definition for $\alpha$, as follows :

%\begin{equation}
%[\alpha]=\frac{\cancel{A}\cancel{s}\cancel{kg}\cancel{m^{2}}\cancel{s^{-2}}\cancel{s}\cancel{kg}\cancel{m^{2}}\cancel{s^{-2}}\cancel{m^{-3}}\cancel{m}}{\cancel{kg^{2}}\cancel{m^{2}}\cancel{s^{-2}}\cancel{A}}=1
%\end{equation}

\begin{acknowledgments}
M. C. Hickey is grateful to the Trinity and the uniformity of nature. We thank the U.S.-U.K. Fulbright Commission for financial support. The work was supported by the ONR (grant no. N00014-09-1-0177), the NSF (grant no. DMR 0504158) and the KIST-MIT program. The authors thank David Cory, Marius Costache and Carlos Egues for helpful discussions.
\end{acknowledgments}

%\bibliography{Gilbert_damping}

\end{document}